\documentclass[%
 reprint,
 amsmath,
 amssymb,
 aps,
 twocolumn,
 superscriptaddress,
 aps,
 prl
]{revtex4-2}

\usepackage{upgreek}
\usepackage{bm}
\usepackage{lipsum}
\usepackage{graphicx,amssymb}
\usepackage{dcolumn}
\usepackage{bm}
\usepackage{xcolor}
\usepackage{amsmath}

\makeatletter
\newcommand*{\rom}[1]{\expandafter\@slowromancap\romannumeral #1@}
\makeatother
\usepackage{amsmath}
\usepackage{graphicx}
\usepackage{bm}
\usepackage[caption=false]{subfig}
\usepackage{float}
\usepackage{placeins}
\usepackage{xcolor}
\usepackage[input-symbols = {text}]{siunitx}
\usepackage{titlesec}
\usepackage{gensymb}
\usepackage{xcolor}
\usepackage{braket}
\usepackage{multirow}
\usepackage{booktabs}
\usepackage{siunitx}
\usepackage{multirow}
\usepackage{booktabs}
\usepackage[
colorlinks=true,
citecolor=blue,
setpagesize=false]{hyperref}

\titleformat{\section}
    {\normalfont\itshape\filcenter}{\thesection}{1em}{}
\titlespacing*{\section}{0pt}{0.5\baselineskip}{\baselineskip}

\begin{document}

\preprint{Corbae \textit{et. al.}}

\title{Probing Coordination Environments in Buried Oxides of Aluminum Josephson Junctions by Resonant X-ray Reflectivity}

\author{Paul Corbae}
\thanks{These authors contributed equally to this work.}
\thanks{Correspondence to: \href{mailto:pjcorbae@slac.stanford.edu}{pjcorbae@slac.stanford.edu}}
\affiliation{Stanford Synchrotron Radiation Lightsource, SLAC National Accelerator Laboratory, 2575 Sand Hill Road, Menlo Park, California 94025, USA}

\author{Alex Abelson}
\thanks{These authors contributed equally to this work.}
\affiliation{Lawrence Livermore National Laboratory, 7000 East Ave, Livermore, California 94550, USA}
\author{Shivani Srivastava}
\affiliation{Lawrence Livermore National Laboratory, 7000 East Ave, Livermore, California 94550, USA}
\author{Heemin Lee}
\affiliation{Department of Physics, Applied Physics, Stanford University, Stanford, California 94305}
\author{Bevin Huang}
\affiliation{Lawrence Livermore National Laboratory, 7000 East Ave, Livermore, California 94550, USA}
\author{Lyrik R-J Lee}
\affiliation{Lawrence Livermore National Laboratory, 7000 East Ave, Livermore, California 94550, USA}
\author{Davis B. Rash}
\affiliation{Lawrence Livermore National Laboratory, 7000 East Ave, Livermore, California 94550, USA}
\author{Cheng-Tai Kuo}
\affiliation{Stanford Synchrotron Radiation Lightsource, SLAC National Accelerator Laboratory, 2575 Sand Hill Road, Menlo Park, California 94025, USA}
\author{Donghui Lu}
\affiliation{Stanford Synchrotron Radiation Lightsource, SLAC National Accelerator Laboratory, 2575 Sand Hill Road, Menlo Park, California 94025, USA}
\author{Mihir Pendharkar}
\affiliation{SLAC National Accelerator Laboratory, 2575 Sand Hill Road, Menlo Park, California 94025, USA}
\affiliation{E.L. Ginzton Laboratory, Stanford University, Stanford, California 94305}
\author{Loren D. Alegria}
\affiliation{Lawrence Livermore National Laboratory, 7000 East Ave, Livermore, California 94550, USA}
\author{Tian T. Li}
\affiliation{Lawrence Livermore National Laboratory, 7000 East Ave, Livermore, California 94550, USA}
\author{Keith G. Ray}
\affiliation{Lawrence Livermore National Laboratory, 7000 East Ave, Livermore, California 94550, USA}
\author{Shannon P. Harvey}
\affiliation{SLAC National Accelerator Laboratory, 2575 Sand Hill Road, Menlo Park, California 94025, USA}
\author{Apurva Mehta}
\affiliation{Linac Coherent Light Source, SLAC National Accelerator Laboratory, 2575 Sand Hill Road, Menlo Park, California 94025, USA}
\author{David I. Schuster}
\affiliation{Department of Applied Physics, Stanford University, Stanford, California 94305}
\affiliation{E.L. Ginzton Laboratory, Stanford University, Stanford, California 94305}
\affiliation{SLAC National Accelerator Laboratory, 2575 Sand Hill Road, Menlo Park, California 94025, USA}
\author{Vincenzo Lordi}
\affiliation{Lawrence Livermore National Laboratory, 7000 East Ave, Livermore, California 94550, USA}
\author{Paul B. Welander}
\affiliation{SLAC National Accelerator Laboratory, 2575 Sand Hill Road, Menlo Park, California 94025, USA}
\author{Jun-Sik Lee}
\thanks{Correspondence to: \href{mailto:jslee@slac.stanford.edu}{jslee@slac.stanford.edu}}
\affiliation{Stanford Synchrotron Radiation Lightsource, SLAC National Accelerator Laboratory, 2575 Sand Hill Road, Menlo Park, California 94025, USA}


\begin{abstract}
Decoherence remains a critical obstacle to achieving high-fidelity, scalable superconducting qubits, with the tunnel barrier of Josephson junctions a key source of loss.
 Here we apply resonant X-ray reflectivity to non-destructively probe the electronic structure of buried layers in Al/AlO$_x$/Al Josephson junctions. At the Al $K$-edge, energy-dependent modulations in the reflectivity maps enable Kramers–Kronig–constrained extraction of the layer-resolved atomic scattering factors. The analysis reveals that the barrier coordination evolves from more tetrahedral toward predominantly octahedral character with increasing oxidation pressure. At the interfaces, the lower metal–oxide boundary is comparatively under-coordinated and disordered relative to the upper interface. Comparison with simulated X-ray absorption spectra identifies the dominant coordination motifs within the oxide and its interfaces, providing depth-resolved structural insight that constrains microscopic models of two-level system formation. These results link growth conditions, local coordination environments, and junction electronic properties, demonstrating resonant X-ray reflectivity as a powerful tool for probing the microscopic materials properties of Josephson junctions and providing a materials-level framework for mitigating decoherence in superconducting qubits.

\end{abstract}

\maketitle

\section*{Introduction}

A central theme in quantum information science (QIS) is the development of qubits with long coherence times and efficient control and read-out \cite{doi:10.1126/science.1231930}. Superconducting qubits incorporating Al/AlO$_x$/Al Josephson junctions have emerged as leading candidates for scalable quantum technologies \cite{Oliver2013,10.1063/1.5089550,annurev:/content/journals/10.1146/annurev-conmatphys-031119-050605}, achieving coherence times over 1 ms in several laboratories \cite{bland2025millisecond, Tuokkola2025}. However, qubit coherence times and uniformity must improve to realize utility-scale quantum computation, with the junction becoming a performance-limiting factor \cite{bland2025millisecond}. Despite the importance of Al/AlO$_x$/Al junctions in many modern superconducting devices, relatively little is known about their atomic scale structure. This gap in understanding motivates quantitative structural and compositional probes that can be applied directly to realistic device stacks. Here, we use resonant X-ray reflectivity to study the structure of buried Al/AlO$_x$/Al junctions and the effect of processing variations on the dominant coordination environments of Al and O within the junction. 

In superconducting qubits, parasitic two-level systems (TLS) are the dominant source of decoherence \cite{Muller_2019, lisenfeld2025mappingpositionstwolevelsystemssurface}. TLS in amorphous materials are hypothesized to be atomic-scale defects in which clusters of atoms tunnel between nearly equivalent configurations in disordered environments \cite{PhysRevApplied.23.024017}. 
The amorphous oxide barrier and its interfaces with the crystalline metal electrodes host a range of atomic coordination environments that may give rise to such configurations, particularly at the metal–oxide interfaces where intrinsically perturbed coordination is expected to support a diverse landscape of local bonding configurations.
Characterizing these local structural motifs is therefore a necessary step toward constraining the microscopic picture of junction-hosted TLS. 

Despite the central role of materials and interfaces in limiting qubit coherence, direct structural insight into Al/AlO$_x$/Al junctions remains limited and is largely derived from transmission electron microscopy (TEM) and related spectroscopies \cite{10.1063/1.4919224,Zeng2016,Fritz2018,Pappas2024}.
TEM provides high spatial resolution and has yielded valuable structural insights, but sample preparation is time-intensive, and the ion milling process
can introduce structural modifications or alter the composition of the Al electrodes \cite{Foxen_2018}. Additionally, because TEM probes a small volume of material, individual measurements may not capture the statistical variation across an entire wafer or fully represent a particular junction growth process. Surface-sensitive techniques such as X-ray photoelectron spectroscopy have been employed successfully to help identify and eliminate lossy native oxides on superconductors within quantum circuits\cite{doi:10.1126/science.abb2823,PhysRevLett.103.197001, PhysRevX.13.041005}. However, the short inelastic mean free path of electrons limits sensitivity to the surface \cite{https://doi.org/10.1002/sia.740010103}.

Resonant X-ray reflectivity (RXR) offers a powerful approach for characterizing buried layers and interfaces by combining the depth sensitivity of non-resonant X-ray reflectivity (XRR) with the elemental and chemical sensitivity of X-ray absorption spectroscopy (XAS) \cite{Hamann-Borrero2017}. By tuning the incident photon energy across the Al and O absorption edges while measuring the specular reflectivity, RXR provides depth-resolved sensitivity to variations in scattering factors associated with distinct chemical environments.

Such depth resolution is particularly critical for Al/AlO$_x$/Al junctions, as Al oxidation follows a Cabrera–Mott mechanism in which Al ions migrate toward the oxygen-rich surface under a chemically driven potential gradient \cite{Cabrera_1949}. Because oxidation ceases at a thickness determined by temperature and oxygen partial pressure, this process naturally generates a compositional and coordination gradient across the oxide, with under-coordinated Al near the internal metal–oxide interface and more fully coordinated sites toward the surface \cite{somjitAtomicElectronicStructure2022,kimDensityfunctionalTheoryStudy2020,tanOxygenStoichiometryInstability2005, Cyster2021}. RXR therefore provides a direct means to resolve how processing parameters influence this buried coordination landscape.

In this work, we use non-destructive, element-sensitive RXR to probe the coordination of buried Al/AlO$_x$/Al junctions prepared under distinct oxidation conditions. Through Kramers–Kronig (KK)-constrained fitting of Al $K$-edge data, we extract layer-resolved atomic scattering factors and, by comparison with \textit{ab initio} X-ray absorption spectra, identify the dominant coordination motifs in the barrier and at each metal–oxide interface. We observe a clear asymmetry between interfaces: the lower metal–oxide interface is more disordered and Al is under-coordinated, whereas the upper interface exhibits a more well-defined, higher-coordination environment. Across a series of junctions grown with increasing oxidation pressure (see Table 1), the barrier coordination evolves from a mixed tetrahedral/octahedral character toward predominantly octahedral coordination. Complementary EELS measurements support these structural and chemical trends, and dc transport measurements provide an electronic perspective, from which nominal junction thicknesses and barrier heights are extracted using tunneling models. For thinner oxides, all three approaches show good agreement, whereas for thicker oxides the effective electronic thickness is smaller than that inferred from reflectivity and EELS. These results demonstrate that a layer-resolved approach of local bonding and coordination environments within the buried tunnel barrier of Josephson junctions provides new constraints on models of two-level system formation, insight into processes governing barrier uniformity, and a pathway toward optimized junction materials for higher-coherence superconducting qubits.

\section*{Methods}

\begin{figure*}
\centering
   \includegraphics[width=0.85\textwidth]{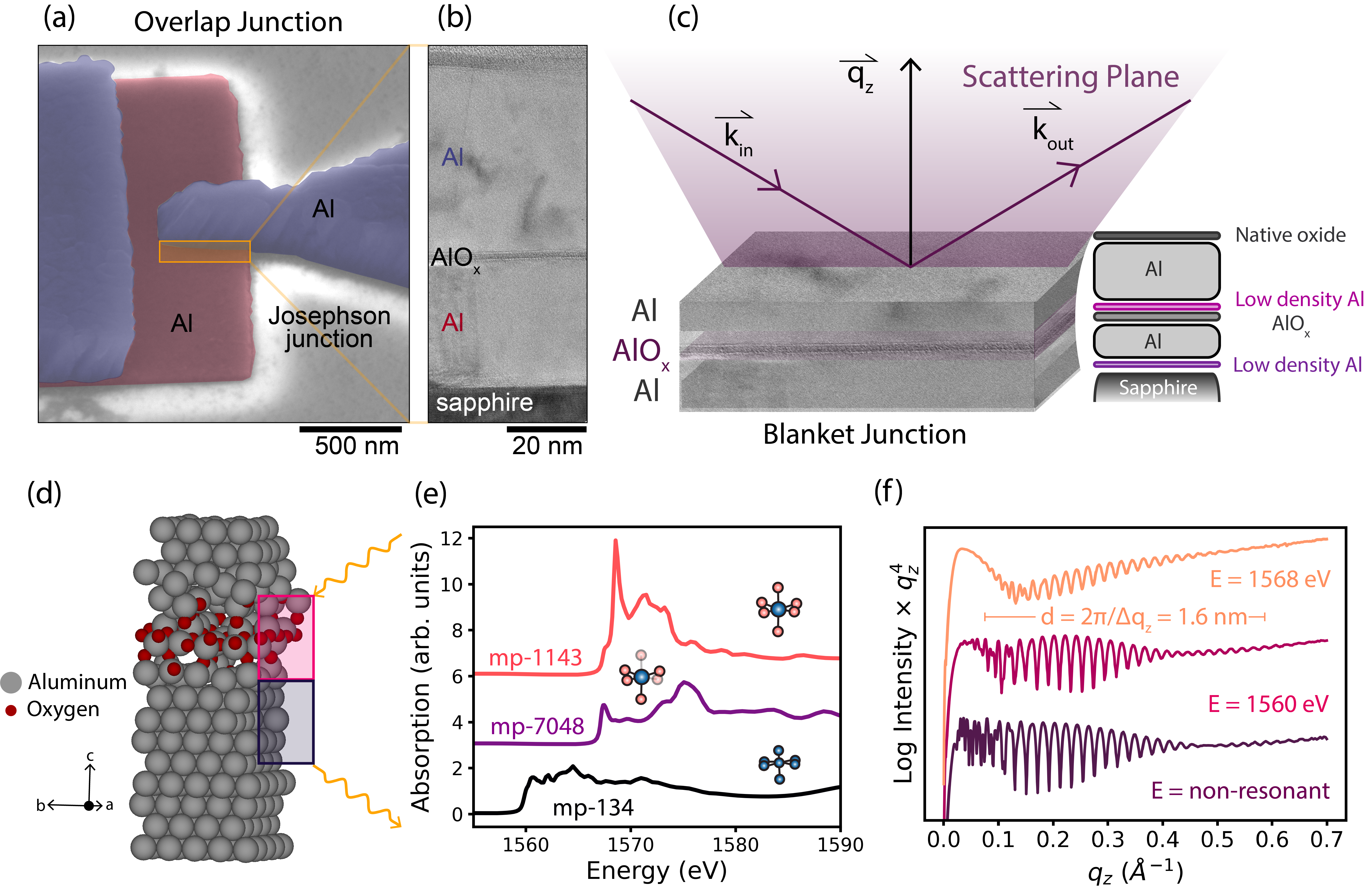}
  \caption{\textbf{Probing buried oxides in Al/AlO$_x$/Al Josephson junctions with resonant X-ray reflectivity.} \textbf{(a)} False-colored SEM image of a typical Josephson junction with the top electrode in purple and base electrode in red. \textbf{(b)} Cross-sectional TEM image of an Al/AlO$_x$/Al junction which clearly resolves the ultrathin AlO$_x$ tunnel barrier between the Al electrodes. \textbf{(c)} Schematic of the resonant X-ray reflectivity (RXR) geometry, where the specular reflectivity is measured under the condition $\theta_{in} = \theta_{out}$. The layer schematic shown in the figure was used for fitting XRR data. See main text for details. \textbf{(d)} Simulated Al/AlO$_x$/Al junction highlighting the amorphous nature of the junction and disorder at the interfaces. \textbf{(e)} Calculated Al $K$-edge X-ray absorption near edge spectra from the Materials Project \cite{Zheng2018,Mathew2018,Horton2025} for metallic Al (mp-134, fcc), monoclinic Al$_2$O$_3$ (mp-7048, C2/m) with mixed tetrahedral and octahedral Al coordination, and $\alpha-$Al$_2$O$_3$ (mp-1143, corundum, R-3c) with purely octahedral Al coordination. The metallic Al spectrum exhibits a broad, featureless near-edge profile, while both oxide phases show sharper absorption features reflecting the localized Al–O bonding environments, with the monoclinic phase displaying additional spectral complexity due to its inequivalent Al sites. \textbf{(f)} $q_z$ line cuts at fixed photon energies demonstrate that the Kiessig fringe periodicity evolves across the Al $K$-edge, with longer-wavelength oscillations at \SI{1568}{eV} (red, purple in (e)) relative to \SI{1560}{eV} (black in (e)), consistent with resonant modulation of the anomalous scattering factors. Rough estimation of the envelope width in $q$ gives a junction thickness $d$ of \SI{1.6}{nm}.}
  \label{fig:sample}
\end{figure*}

We fabricated a series of Al/AlO$_x$/Al junctions on \textit{c}-plane sapphire using double-angled electron beam evaporation (see Supplemental Information). X-ray studies used blanket films while dc transport measurements used small-area junctions (see Fig. \ref{fig:sample}(a-c)) patterned with electron beam lithography. The Al/AlO$_x$/Al junctions consist of a 30 nm Al layer, an oxide layer produced according to various conditions noted in Table \ref{tab:oxidation_conditions} at room temperature, and a 50 nm Al cap (see Fig. ~\ref{fig:sample}(b). Additional fabrication details are provided in the Methods section of the Supplemental Material.

\begin{table}[t]
\centering
\caption{Junction growth conditions}
\label{tab:oxidation_conditions}

\begin{tabular}{p{0.18\columnwidth} p{0.57\columnwidth} p{0.22\columnwidth}}
\toprule
Sample ID & Oxidation condition & Junction thickness ($d$) \\
\midrule
~~~~~A & 15 minutes at 4.5 Torr O$_2$ & ~~~~0.8 nm \\
~~~~~B & 15 minutes at 45 Torr O$_2$ & ~~~~1.7 nm \\
~~~~~C & 15 minutes at 45 Torr O$_2$, then 3 rounds of 1 nm Al  at 0.1 nm$\cdot$s$^{-1}$, 15 minutes at 45 Torr O$_2$  & ~~~~4.8 nm \\
\bottomrule
\end{tabular}
\end{table}

\begin{figure*}
\centering
  \includegraphics[width=0.93\textwidth]{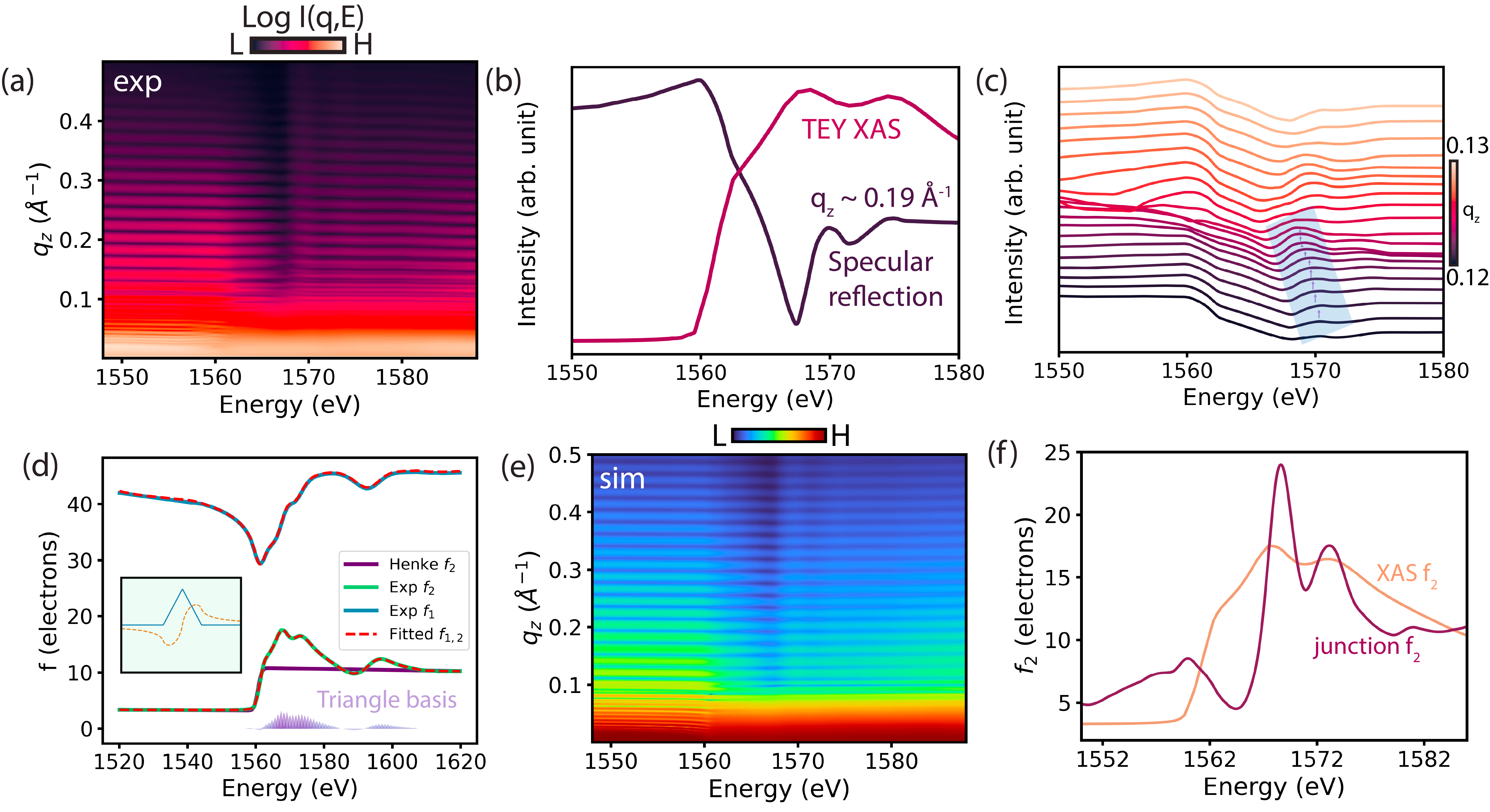}
  \caption{\textbf{Extracting the electronic structure in buried Al/AlO$_x$/Al Josephson junctions via variational fitting of RXR maps.} 
  \textbf{(a)} Al $K$-edge RXR map for sample B, showing reflectivity attenuation near $\sim$ \SI{1568}{} and \SI{1574}{eV} that fingerprints the local electronic environment. The colorbar ranges from low (L) to high (H) intensity. \textbf{(b)} An energy-dependent line cut at $q_z = \SI{0.19}{\AA^{-1}}$ reveals a pronounced dip in specular reflectivity near $\sim$ \SI{1570}{eV} arising from absorption driven by the imaginary scattering factor $f_2$. Al $K$-edge TEY XAS spectrum exhibiting lineshape characteristics predominantly consistent with metallic aluminum. The dip lines up with the peak in XAS. \textbf{(c)} Energy-dependent line cuts at fixed $q_z$ reveal a pronounced peak in specular reflectivity near $\sim$ \SI{1570}{eV} shifting in energy with $q_z$, arising from constructive interference driven by modulation of the real scattering factor $f_1$. \textbf{(d)} Atomic scattering factors $f_{1,2}$ for Al$_2$O$_3$ derived from experimental TEY XAS; triangle basis functions used to parameterize $f_2$ are also shown in the inset. The atomic scattering factors are expressed in units of electrons because they quantify the scattering strength of a multi-electron system relative to that of a single free electron. \textbf{(e)} Simulated RXR map following variational refinement, capturing the reflectivity suppression at \SI{1568}{eV} and \SI{1574}{eV}. \textbf{(f)} Refined $f_2$ spectrum, compared to the $f_2$ derived from XAS, for the buried junction layer obtained from variational fitting, revealing distinct spectral features at energies corresponding to the observed reflectivity minima corresponding to an octahedral coordination environment.
  }
  \label{fig:structure}
\end{figure*}

Structural and spectroscopic characterization was carried out using RXR at Beamline 13-3 of the Stanford Synchrotron Radiation Lightsource (SSRL) at SLAC. Figure~\ref{fig:sample}(c) shows the experimental geometry, in which the reflected intensity is measured under the specular condition as a function of the out-of-plane momentum transfer ($q_z = 4\pi \sin\theta /\lambda$ where $\lambda$ is wavelength and $\theta$ is incidence angle). XRR encompasses both resonant and non-resonant measurements, whereas RXR refers specifically to measurements performed at photon energies near an absorption edge. To enhance contrast between the Al electrodes and the AlO$_x$ junction, we used $\sigma$-polarized photons with energies spanning \SIrange{1540}{1590}{eV}, corresponding to the Al $K$-edge. The observed Kiessig fringes arise from interference between waves reflected at interfaces separating layers with different refractive indices. For X-rays, the refractive index depends on the complex atomic scattering factors $f_1$ and $f_2$, 
\begin{equation}
    n(\omega) = 1 - \frac{n_a r_e \lambda^2}{2\pi}\left[f_1(\omega) - i f_2(\omega)\right],
\end{equation}
where $n_a$ is the number density, and $r_e$ the classical electron radius. Far from absorption edges, $f_{1,2}$ exhibit only weak energy dependence; however, near a resonance, they vary strongly due to changes from the local electronic environment. 
The imaginary part of the atomic scattering factor, $f_2$, is directly related to the X-ray absorption spectrum. The XAS spectra, acquired in total electron yield (TEY) mode, were converted to $f_2$ and KK transformed to obtain $f_1$ for resonant reflectivity modeling \cite{Watts:14}.

At photon energies well away from the absorption edges, tabulated values of $f_{1,2}$ were used \cite{HENKE1993181}. The underlying layer structure was first determined by fitting reflectivity scans collected at non-resonant photon energies of \SI{1548}{} and \SI{524}{eV} for each growth condition \cite{Glavic:ge5118}. The resulting structural model, illustrated in the inset of Fig.~\ref{fig:sample}(c), includes the substrate, bottom electrode, junction oxide, top electrode, native oxide, and two thin interfacial layers introduced to account for reduced electron density (structural parameters presented in Supplemental Information). The layer at the substrate–Al interface was introduced to improve fit quality, consistent with prior reports \cite{https://doi.org/10.1002/advs.202413058}, while the layer above the junction derives from first-principles calculations, which predict thin regions of reduced Al density arising from the incommensurate lattice constants of Al and AlO$_x$ (see Supplemental Information for details).
This non-resonant refinement establishes the structural framework used in subsequent resonant (i.e., RXR) analysis.

\section*{Results}

Tuning the photon energy to an absorption edge in a reflectivity measurement (RXR) encodes the local electronic and chemical environment of each layer into the measured signal through the energy-dependent atomic scattering factors $f$. This is particularly relevant for amorphous AlO$_x$ barriers [Fig.~\ref{fig:sample}(d)], which host a distribution of coordination environments. As shown in Fig.~\ref{fig:sample}(e), XAS spectra exhibit distinct features associated with metallic Al–Al bonding and Al–O coordination \cite{Altman2017}, providing spectroscopic fingerprints of the different local environments present within the junction. RXR therefore extends the structural information obtained from non-resonant XRR by adding sensitivity to coordination-dependent variations in bonding.

Based on these spectral features, we measure reflectivity at \SI{1560}{eV}, corresponding to metallic Al–Al absorption, and at \SI{1568}{eV}, associated with a sharp Al–O bonding resonance. Representative scans are shown in Fig.~\ref{fig:sample}(f), with the intensity multiplied by $q_z^4$ to compensate for the Fresnel decay. At the AlO$_x$ resonance ($\sim\SI{1568}{eV}$), the reflectivity profile exhibits a low-frequency component with a characteristic length of \SI{1.6}{nm}, consistent with the tunnel barrier thickness, superimposed on the high-frequency oscillations present in the non-resonant scans. In contrast, tuning to the metallic Al resonance ($\sim\SI{1560}{eV}$) produces a profile similar to the non-resonant case. This differential response demonstrates how RXR selectively amplifies the scattering contrast of specific layers. 

We extract the complex scattering factors (i.e., $f_1$ and $f_2$) of the buried oxide layer to reconstruct its layer-resolved near-edge spectral response by extending the measurement across the full Al $K$-edge. The resulting RXR map is shown in Fig.~\ref{fig:structure}(a), where vertical cuts reproduce conventional XRR profiles as a function of $q_z$, while horizontal cuts trace the energy dependence of the specular reflectivity at fixed $q_z$. The former primarily reflects the structural layering of the junction, whereas the latter captures the resonant modulation arising from the energy-dependent dispersion ($f_1$) and absorption ($f_2$) of each layer. Two prominent features are evident in this two-dimensional RXR map. First, a clear intensity minimum appears near \SI{1568}{eV}, corresponding to the strong absorption resonance at this energy, as seen in Fig.~\ref{fig:structure}(b). We associate this feature with absorption by 6-fold coordinated Al–O environments, consistent with the XAS spectrum in Fig.~\ref{fig:sample}(e). Second, subtle fringe crossings emerge at low $q_z$ near the same energy. These arise from energy-dependent phase shifts in the reflected waves due to enhanced contrast at resonance. The effect is more clearly resolved in the horizontal linecuts shown in Fig.~\ref{fig:structure}(c), superimposed on the broader $q_z$ modulation discussed above.

We determine the complex scattering factors of the buried oxide through variational fitting of the RXR maps \cite{PhysRevB.86.024102,10.1063/1.1979470,10.1063/5.0187303,DVORTSOVA2024110456,DVORTSOVA2024140517}. This process reconstructs the near-edge spectral response of each layer within the structure, with direct sensitivity to the local coordination environment. Each layer in the reflectivity model is initialized using structural parameters obtained from the non-resonant XRR fits, including thickness, density, and roughness. The energy dependence of each layer is described by a complex atomic scattering factor, $f(\omega)=f_1(\omega)-if_2(\omega)$. For the metallic Al layers, $f(\omega)$ is taken directly from the converted XAS spectrum shown in Fig.~\ref{fig:structure}(b), which is dominated by $1s\rightarrow3p-\pi^b$ (\SI{1562.5}~{eV}), $1s\rightarrow3p-\sigma^*$ (\SI{1567.5}~{eV}), and $1s\rightarrow3d-\sigma^*$ (\SI{1573.5}~{eV}) transitions characteristic of bulk Al; these spectral features reflect the large volume fraction of metallic Al within the limited TEY sampling depth.

For the AlO$_x$ layer, the scattering factor is constructed as $f = 2f_{\rm{Al}} + 3f_{\rm{O}}$, with variational refinement of the spectral shape and magnitude to accommodate deviations in stoichiometry $x$. Deviations of each layer’s $f_2(\omega)$ from the tabulated Henke values are parameterized using triangle basis functions following work by Stone \textit{et al.} \cite{PhysRevB.86.024102}. These basis functions possess analytic KK transforms, ensuring that the dispersive component $f_1(\omega)$ remains self-consistent (see Fig.~\ref{fig:structure}(d)). By fitting only the difference relative to the tabulated values, the scattering factors recover the correct asymptotic behavior far from resonance while preserving the validity of the KK relation. The resulting variational refinement reproduces the RXR map shown in Fig.~\ref{fig:structure}(e).

\begin{figure*}
\centering
  \includegraphics[width=0.97\textwidth]{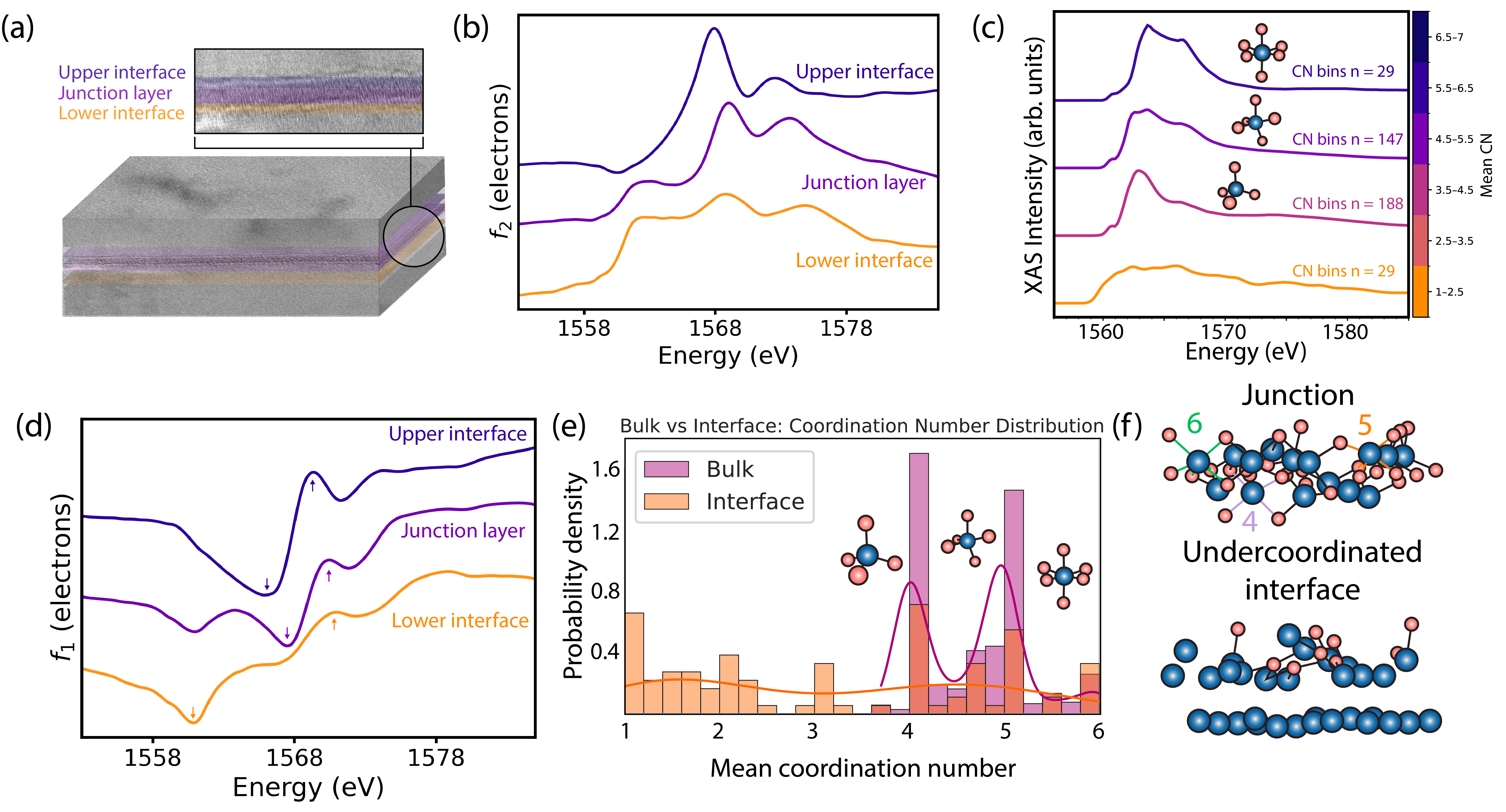}
  \caption{\textbf{Resolving the buried interface electronic structure in Al/AlO$_x$/Al Josephson junctions.} \textbf{(a)} Interface structure of the AlO$_x$ barrier layer from TEM. The lower metal–oxide interface forms by oxidation of the base Al layer and is characteristically oxygen-deficient, while the upper interface is formed during Al deposition onto the completed oxide and results in structurally and chemically distinct interfaces. 
  \textbf{(b)} $f_2(E)$ for the lower interface, the junction, and the upper interface. The upper interface shows a dominant 6-fold coordination environment while the lower interface shows less apparent structure due to structural disorder and higher Al content. \textbf{(c)} Simulated XAS spectra for Al with differing coordination number. As the coordination increases the main peak shifts towards higher energy and splits into two.\textbf{(d)} $f_1(E)$ for the lower interface, the junction, and the upper interface (same color map as b). The upper interface shows a high effective electron density near \SI{1568}{eV} where the lower interface shows low effective electron density near \SI{1560}{eV} as large variations in $f_2$ drive correspondingly large changes in $f_1$ through KK relations. Curves are offset by a constant amount for visual clarity. \textbf{(e)} Coordination environment analysis for bulk and interfacial AlO$_x$ in simulated Josephson junctions. The interface exhibits a broad distribution of local coordination environments spanning 1- to 6-fold, whereas the bulk is confined predominantly to 4- to 6-fold sites. The curves are the probability distribution function from the histograms. \textbf{(f)} First-principles derived snapshots of different regions in the AlO$_x$ junction. Tetrahedral, trigonal bipyramidal, octahedral (4,5,6) environments are highlighted.
  }
  \label{fig:interface}
\end{figure*}

Converged fits to the RXR maps were obtained by allowing variation in $f(\omega)$ functions of the junction oxide, native oxide, and substrate layers, as the absorption from AlO$_x$ accounts for many of the observed features of the RXR maps. Allowing the atomic scattering factors of the low-density Al layer to vary slightly improves the goodness of fit but does not change the the junction oxide's $f_2$, so we hold its scattering factors fixed to reduce the number of fit parameters. The resulting $f_2(\omega)$ for the buried junction layer is shown in Fig. \ref{fig:structure}(f).  Comparison with both XAS spectra of bulk $\alpha$-Al$_2$O$_3$ and simulation (see below) of higher-coordinated amorphous AlO$_x$ ($CN=5,6$) reveals that the variationally extracted $f_2$ is consistent with predominantly octahedral coordination, exhibiting the characteristic sharp, two-peak structure [$1s\rightarrow3p\sigma/\pi^{*} (t_{1u}$) and $1s\rightarrow3d\sigma^{*}(t_{2g}$)] at higher energy seen in Fig. \ref{fig:sample}(e) \cite{Altman2017}.

\begin{figure*}
\centering
  \includegraphics[width=1\textwidth]{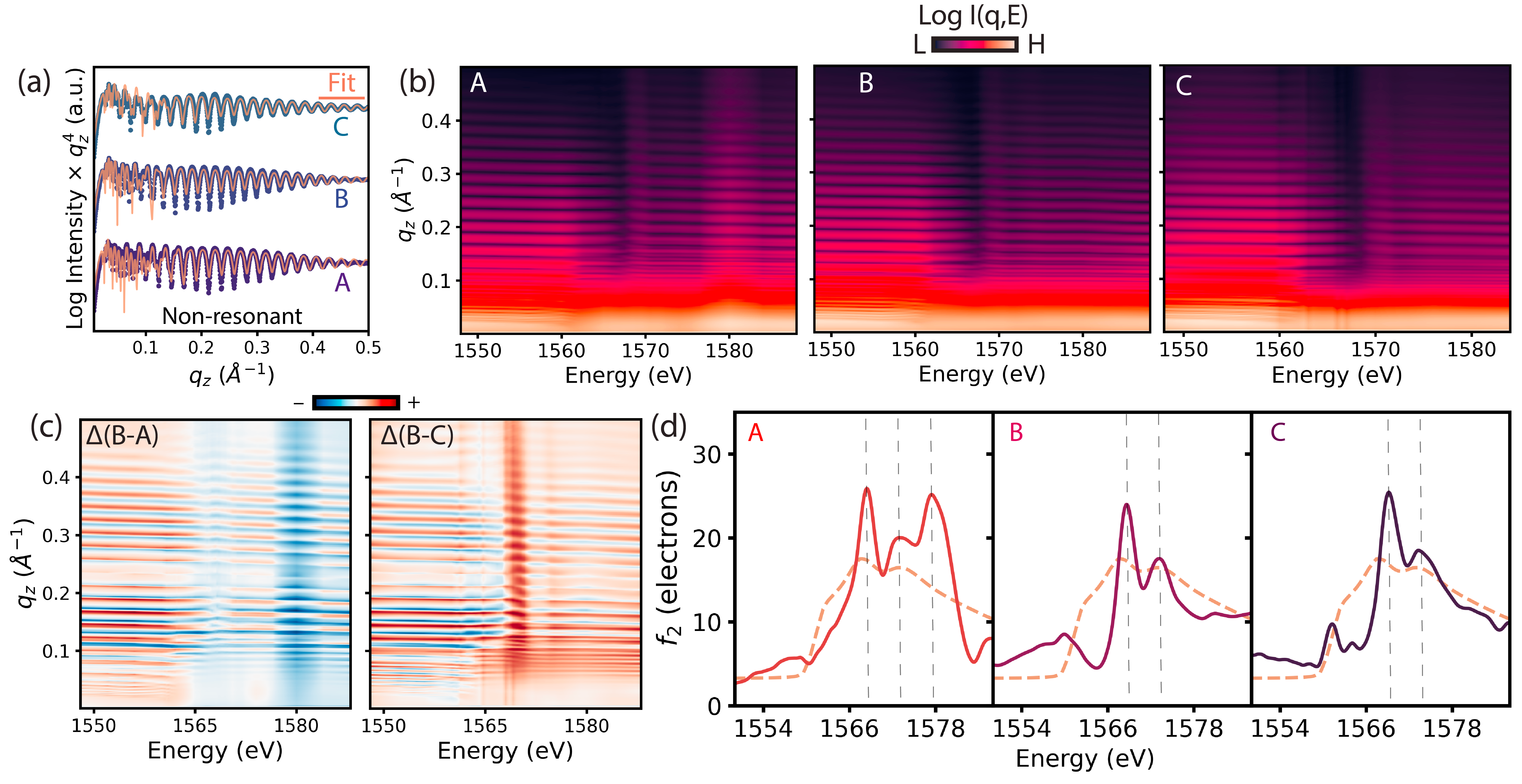}
  \caption{\textbf{Growth-dependent electronic structure of Al/AlO$_x$/Al Josephson junctions probed by Al $K$-edge RXR.} \textbf{(a)} Non-resonant XRR measured below the Al $K$-edge with the corresponding structural fit for sample A, B, and C. \textbf{(b)} Al $K$-edge RXR maps across the junction growth series. With increasing O$_2$ pressure, the reflectivity exhibits enhanced attenuation near $\sim$\SI{1568}{eV} and a suppression of the feature around $\sim$\SI{1577}{eV}, indicating systematic changes in the local electronic environment. \textbf{(c)} Difference maps between samples B - A as well and B - C highlight pronounced energy-dependent modulations, reflecting variations in the electronic structure and coordination of the buried oxide layer as well as structural parameters. \textbf{(d)} The extracted $f_2(E)$ of the buried junction layer for A, B, and C as well as the initial condition $f_2(E)$ extracted from the XAS. The $f_2(E)$ moves from having three main peaks in A to two main peaks in B/C representing a change in local coordination from a mixture of tetrahedral and octahedral to mostly octahedral.}
  \label{fig:theory}
\end{figure*}

While the extracted $f_2(\omega)$ reflects the average coordination within the junction oxide, the layer-resolved refinement further enables us to distinguish the electronic structure of the metal–oxide interfaces from that of the interior oxide. To this end, we isolate the interfacial regions by removing \SI{3}{\AA} from either side of the junction oxide and treat these slices as distinct layers whose complex scattering factors are refined independently [Fig.~\ref{fig:interface}(a)]. As shown in Figure~\ref{fig:interface}(b), the imaginary component $f_2$ at the lower interface exhibits markedly reduced spectral structure across the Al $K$-edge, indicative of a disordered and under-coordinated local environment. In contrast, the interior oxide retains well-defined spectral features characteristic of mixed 4- to 6-fold Al coordination, albeit with reduced relative intensity compared to fully octahedral Al$_2$O$_3$, consistent with a distorted or partially coordinated environment. The upper interface displays $f_2$ line shapes that more closely resemble those of a 6-fold coordinated oxide, suggesting a comparatively well-ordered and densely coordinated interfacial layer.

These trends are consistent with first-principles simulations of oxide growth in Al-based Josephson junctions \cite{Cyster2021}, which predict asymmetric oxygen distribution across the barrier. In those simulations, oxygen preferentially migrates toward the top electrode during growth, leading to a denser oxide at the upper interface and a more defective, under-coordinated environment at the lower interface.

To directly link RXR-derived junction $f$ with structural models of Al/AlO$_x$/Al junctions, we performed density functional theory (DFT) calculations and computed XAS spectra for different local coordination environments [Fig. \ref{fig:interface}(c)]. Previous DFT studies have shown that AlO$_x$ exhibits a thermodynamic preference for amorphous films over the $\alpha$ and $\gamma$ crystalline phases at thicknesses below \SI{1}{nm} \cite{tanOxygenStoichiometryInstability2005,reichelAmorphousCrystallineState2008}. We employ thin interface models with non-ideal stoichiometry at the Al/AlO$_x$ interface and well as bulk models of stoichiometric Al$_2$O$_3$ obtained from the Materials Project database \cite{jainCommentaryMaterialsProject2013} to calculate theoretical X-ray absorption spectra using DFT. Continuous symmetry measure (CSM) analysis \cite{waroquiersStatisticalAnalysisCoordination2017} is performed to treat distorted Al-O coordination environments and to smoothly interpolate between nominal integer coordination numbers (see Supplemental Information).

The calculated spectra reveal systematic trends with coordination number. The main absorption peak shifts to higher energy as the coordination increases, highlighting the sensitivity of the near-edge response to local bonding geometry.
A pronounced pre-edge or shoulder feature appears for mean Al coordination numbers of approximately 3.5-4.5, with additional weaker contributions arising from distorted 6-fold environments. For higher-coordination environments (5- to 6-fold), a broader secondary peak-like feature emerges at higher energies. These spectral variations reflect changes in Al $3s$ - O $3p$ hybridization associated with reduced symmetry, enabling transitions absent in the undistorted octahedral environments of crystalline $\alpha$-Al$_2$O$_3$. The coordination-number distribution also varies with local chemistry: interface regions contain a substantial population of under-coordinated Al sites (absent in stoichiometric amorphous Al$_2$O$_3$), giving rise to the distinct spectral fingerprints that we observe. 

This interpretation is further corroborated by the behavior of $f_1$ across the junction, shown in Fig.~\ref{fig:interface}(d). In the oscillator model, $f_1(\omega)$ approaches $Z$ in the high-frequency limit of $\omega^2 \gg \omega_s^2$, where $\omega_s$ is the resonance frequency. Near an absorption edge, the real part of the atomic scattering factor devisates sharply from $Z$, and the quantity $f_1(\omega) = Z + \Delta f_1(\omega)$ can be interpreted as an effective scattering charge that depends on photon energy. Within this framework, the lower interface exhibits reduced effective electron density at photon energies corresponding to metallic Al and an increase at photon energies associated with AlO$_x$. Taken together, the trends in the DFT and RXR-derived $f_2$ indicate that the bulk of the junction in this sample (sample B) is predominantly octahedrally coordinated, though with possible contributions from near-octahedral sites (5-fold). Additionally, there is a redistribution of electronic charge and local bonding character at the lower interface, consistent with an oxygen-deficient, metal-proximal environment that is electronically distinct from both the bulk junction oxide and the more fully coordinated upper interface, shown in Fig. \ref{fig:interface}(e,f), in agreement with previous computational and experimental studies \cite{somjitAtomicElectronicStructure2022,kimDensityfunctionalTheoryStudy2020,tanOxygenStoichiometryInstability2005}. 
This interfacial asymmetry underscores that the buried barrier cannot be regarded as structurally uniform, even at sub-nanometer thicknesses.

Having established the coordination environment of the junction oxide, we now examine how it evolves under different oxidation conditions. Correlating the coordination environment with device performance metrics offers a pathway toward identifying optimal fabrication conditions and elucidating the structural origins of decoherence in superconducting qubits. Figure~\ref{fig:theory}(a) shows the non-resonant XRR data for each of the samples within the growth series. Comparing the three sample types, we observe a monotonic increase in the junction thickness from \SI{0.8}{} to \SI{1.8}{} to \SI{4.75}{nm} for A, B, and C, respectively, as the oxidation conditions become progressively more aggressive. 

Figure~\ref{fig:theory}(b) shows resonant XRR maps from three samples within the junction growth series (A, B, and C). There are subtle variations in the specular reflectivity across the series, reflecting distinct electronic environments in the buried junction layers of each film. Additionally, Sample A shows an intensity peak around \SI{1580}{~eV}, indicative of redistributed spectral weight at higher energies. A difference map between Samples A and B RXR maps [Fig. \ref{fig:theory}(c)] highlights differences both structurally in the $q_z$ direction and electronically in the $E$ direction. To fully register the differences between sample types, we perform variational fitting of the other two samples.

Figure \ref{fig:theory}(d) shows the complex parts of the scattering factors for the buried oxide layer of each of the three samples. The $f_2$ spectra exhibit clear differences across growth conditions. In particular, $f_2$ evolves from three distinct peaks at \SI{1568.5}{~eV}, \SI{1573.0}{~eV}, and \SI{1577.7}{~eV} in Sample A to two peaks at \SI{1568.5}{~eV} and \SI{1573}{~eV} in Samples B, C. The $f_2$ spectra for Samples B and C closely resemble the XAS spectrum of 6-fold coordinated AlO$_x$ and are consistent with the DFT [Fig. \ref{fig:interface}(c)]. In contrast, Sample A exhibits an additional higher-energy feature reminiscent of the third peak in $\gamma$-Al$_2$O$_3$ (attributed to Al $1s$ transitions into final states with Al $3d$ tetrahedral symmetry), suggesting an increased fraction of tetrahedral local coordination compared to octahedral \cite{Altman2017,A904286E,Cyster2021}. These systematic spectral changes reveal that increasing oxidation pressure shifts the coordination landscape of the barrier from a mixed tetrahedral/octahedral environment toward predominantly octahedral coordination.

\begin{figure*}
\centering
  \includegraphics[width=1\textwidth]{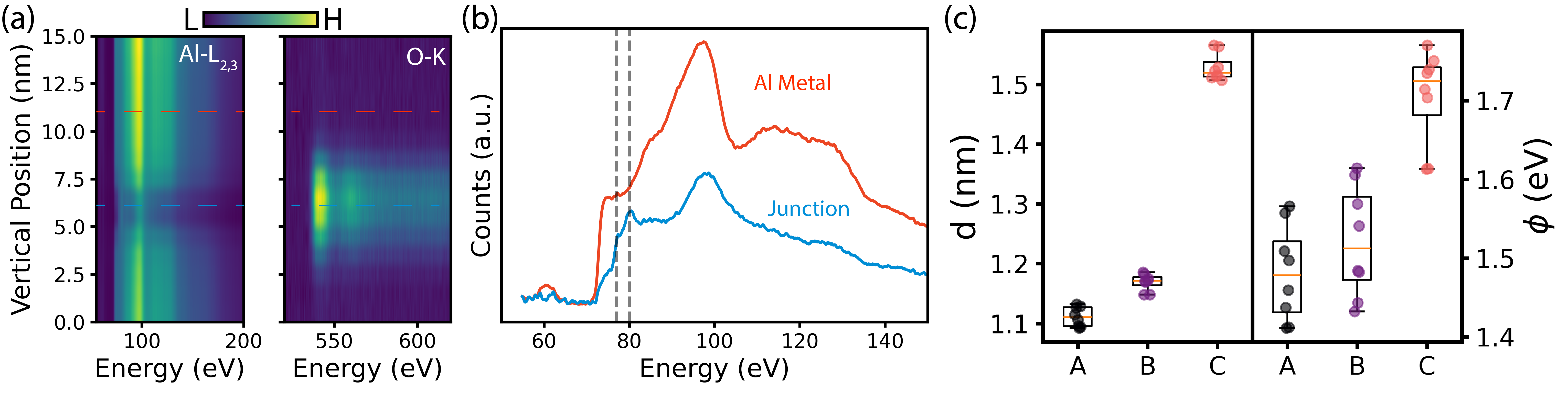}
  \caption{\textbf{Benchmarking RXR with ELNES and IV characterization of Al/AlO$_x$/Al Josephson junctions.} \textbf{(a)} Al $L_{2,3}$-edge of B showing clear intensity change across the junction oxide, with either side displaying metallic aluminum line shapes. O $K$-edge of sample B showing clear O intensity in the junction, exactly where the Al $L_{2,3}$-edge shows a decrease in intensity. \textbf{(b)} Specific Al $L_{2,3}$-edge line cuts in the junction and the Al metal. The JJ oxide has peaks at $\sim$\SI{77}{eV} and \SI{80}{eV} whose ratio determines the ratio of tetrahedral to octahedral environments. 
  \textbf{(c)} Simmons model fits to the measured current–voltage characteristics for the junction series. Both the extracted barrier thickness, $d$, and barrier height, $\phi$, increase with increasing oxidation pressure, consistent with the growth of a thicker, more fully oxidized tunnel barrier.}
  \label{fig:EELS}
\end{figure*}

As an independent probe of local structure, we performed EELS measurements at the Al $L_{2,3}$ and O $K$ edges. Energy loss near-edge structure (ELNES) for Sample B is shown in Fig.~\ref{fig:EELS}.
Clear contrast of the junction is seen from both edges. At the Al $L_{2,3}$ edge, the spectral features at \SI{78}{eV} and \SI{80}{eV} correspond to tetrahedral and octahedral coordination, respectively \cite{10.1063/1.1629397,Filatova2015}. Sample B exhibits ELNES spectra characteristic of predominantly octahedral Al coordination in the oxide, with a high octahedral-to-tetrahedral peak ratio. The O $K$ edge ELNES is less definitive in distinguishing tetrahedral from octahedral coordination \cite{10.1063/1.4793473,Filatova2017,Fritz2018,Filatova2015}. While ELNES provides qualitative insight into coordination environments through relative peak intensities, Al $K$-edge X-ray spectroscopy exhibits well-resolved, coordination-specific spectral features enabling more definitive identification of local bonding configurations \cite{Pappas2024,10.1063/1.1629397,Filatova2015,10.1063/1.4793473,Filatova2017,Fritz2018}. 

Finally, to directly probe the electronic properties of each sample type, we patterned shadow-evaporated Josephson junctions [Fig. \ref{fig:sample}(a)] and performed current-voltage ($I-V$) measurements.  Junction barrier height and thickness were extracted by fitting  $I-V$ data to the Simmons tunneling model \cite{simmonsth}, as shown in Fig. \ref{fig:EELS}(c). The extracted mean thickness increases from \SI{1.12}{nm} and \SI{1.18}{nm} for Samples A and B, respectively, to \SI{1.52}{nm} for Sample C. The significant disparity in RXR and Simmons model-derived thickness for Sample C is likely due to the combination of increased junction roughness and the exponential dependence of tunneling current on thickness: thinner regions disproportionately dominate the electrical response. The extracted barrier heights in Samples A, B and C increase from \SI{1.48}{eV} to \SI{1.52}{eV} and \SI{1.73}{eV}. This trend is consistent with improved oxidation and a reduction in defect-assisted tunneling and a more insulating, bulk-like oxide.

\section*{Discussion}

The evolution from a three- to two-peak profile in $f_2$ across samples A–C, indicates that the amorphous AlO$_x$ barrier becomes increasingly octahedrally coordinated with oxidation pressure, while sample A retains a larger fraction of tetrahedrally coordinated Al sites. This trend is consistent with our finding that the lower metal/oxide interface is under-coordinated: for the thinnest barrier/lowest oxidation pressure (Sample A), a greater proportion of the oxide volume lies in proximity to this interface, shifting the coordination toward lower values.
It is particularly noteworthy that Samples B and C exhibit similar scattering factors given their distinct growth conditions, although they share a high oxygen pressure (45 Torr), suggesting that coordination environment may be more sensitive to oxygen partial pressure during oxidation than total film thickness. 
The O $K$-edge variational fits further corroborate this picture, showing that the dominant peak in $f_2$ is shifted toward lower photon energies for Samples B and C, consistent with a more 6-fold–coordinated local environment \cite{Altman2017} (see Supplemental Information).
Differences in interface structures and variation of interface participation in junction properties may vary with thickness, for typical oxidation conditions.

There are apparent pre-peak features in the extracted $f_2$ of the buried junction layers at \SI{1562}{eV} and \SI{1564}{eV}. The pre-peak at \SI{1562}{eV} can be attributed to a trigonal bipyramid coordination (5-fold) environment, which can be expected in an amorphous layer \cite{D2SC04035B}. The pre-peak observed at \SI{1564}{eV} ($1s\rightarrow2s\sigma^{*} (a_{1g})$) has been ascribed to thermally induced lattice vibrations that amplify octahedral distortions, thereby facilitating enhanced $sp$ hybridization at the lower edge of the conduction band \cite{PhysRevB.85.224108,PhysRevB.111.165112,CABARET199621}. The extent to which the variational fitting technique is sufficiently sensitive to resolve such subtle pre-peak modifications in the electronic structure remains uncertain and should be rigorously validated through its application to a broader range of systems, though this pre-peak feature is captured in our first principles result, Fig. \ref{fig:interface}(c).

Previous studies of the near-edge features of the EELS spectra combined with a pair distribution function analysis from nanobeam diffraction data indicate tetrahedrally coordinated Al atoms in the barrier interior, consistent with amorphous aluminum oxide, but less coordination near the interface, indicative of oxygen vacancies \cite{Zeng2016}. It has been shown with ELNES that atomic-layer deposited (ALD) Al$_2$O$_3$ contains both tetrahedral and octahedral coordination environments, with octahedral Al dominating away from the interface (deeper in the junction), while the interfacial region is comparatively under-coordinated and richer in tetrahedral Al\cite{10.1063/1.1629397}. These results for ALD  Al$_2$O$_3$ are consistent with our findings that in the thicker junctions grown with higher O$_2$ pressure the environment is more octahedral, whereas in the thinner junctions it looks more tetrahedral, Fig. \ref{fig:sample}(d),\ref{fig:theory}(d). It has been demonstrated that replacing amorphous AlO$_x$ with a single-crystal epitaxial Al$_2$O$_3$ tunnel barrier, with an octahedral environment, in Josephson junction qubits reduces the density of decoherence-inducing two-level fluctuators by $\sim 80\%$ \cite{PhysRevB.74.100502}. 
These results indicate that oxidation conditions systematically influence the local coordination landscape of the barrier, thereby constraining the types of defect configurations implicated in TLS formation. Establishing a direct link to qubit coherence, however, will require further device-level investigation.

\section*{Conclusion}

This work represents a direct application of X-ray scattering and spectroscopy to non-destructively resolve the coordination environment and microscopic structure of the buried oxide barrier in Al-based Josephson junctions.
Using Kramers–Kronig-constrained variational fitting of resonant X-ray reflectivity maps across the Al $K$-edge, we  extracted the layer-resolved atomic scattering factors of the buried AlO$_x$ barrier and, through comparison with \textit{ab initio} X-ray absorption spectra, identified the dominant coordination motifs in the bulk oxide and at each metal–oxide interface. Across a series of junctions grown under varying oxidation conditions, we find that the barrier coordination evolves from more tetrahedral toward predominantly octahedral character with increasing oxidation pressure, while the lower metal–oxide interface remains consistently more disordered and under-coordinated than the upper interface.  EELS and $I-V$ measurements independently confirm these structural trends and connect the structure variations to differences in junction electrical characteristics (barrier height). Through systematic variation of growth parameters, we demonstrate how oxidation conditions dictate the local coordination environment and, by extension, constrain the phase space of structural defects implicated in TLS formation as well.
Looking forward, RXR opens opportunities to explore more ambitious process variations: alternative barrier materials, novel oxidation schemes, and epitaxial interfaces with the spectroscopic precision necessary to guide rational optimization, with relatively high throughput and large sampling area. Thus, RXR offers a powerful tool toward the informed, rational design of higher-coherence superconducting qubits.

\section*{Acknowledgments}
We thank Kevin Stone for helpful discussion regarding variational fitting and Cheng Peng for help running FEFF calculation. 
The RXR experiments were carried out at the SSRL (beamline 13-3), SLAC National Accelerator Laboratory, supported by the U.S. DOE, Office of Science, Office of Basic Energy Sciences under Contract No. DE-AC02-76SF00515.
This work was also supported by the auspices of the U.S. Department of Energy by Lawrence Livermore National Laboratory under Contract DE‑AC52‑07NA27344. 
The measured junctions are based upon work supported by the Air Force Office of Scientific Research under award number FA9550-23-1-0692.

\section*{Author contributions}
P.C., A.A., S.S., A.M., P.W., V.L., and J.S.L. conceived the project. P.C., H.L., C.T.K., and J.S.L. performed the XRR measurements. A.A., D.R., B.H., L.R.L. performed the growth, fabrication, and transport. S.S. performed the DFT calculations. 
All authors contributed in discussing the results and forming conclusions.

\bibliographystyle{apsrev4-2}
\bibliography{bib.bib}

\end{document}